# Curvilinear object segmentation in medical images based on ODoS filter and deep learning network

Yuanyuan Peng[1,2,*], Lin Pan[1], Pengpeng Luan[1], Hongbin Tu[1], and Xiong Li[3]

[1]School of Electrical and Automation Engineering, East China Jiaotong University, Nanchang 330000, China
[2]School of Computer Science, Northwestern Polytechnical University, Xi'An 710000, China
[3]School of Software, East China Jiaotong University, NanChang 330000, China

*Corresponding author(s). Email(s): pengmi467347713@126.com;

**Abstract:** Automatic segmentation of curvilinear objects in medical images plays an important role in the diagnosis and evaluation of human diseases, yet it is a challenging uncertainty in the complex segmentation tasks due to different issues such as various image appearances, low contrast between curvilinear objects and their surrounding backgrounds, thin and uneven curvilinear structures, and improper background illumination conditions. To overcome these challenges, we present a unique curvilinear structure segmentation framework based on an oriented derivative of stick (ODoS) filter and a deep learning network for curvilinear object segmentation in medical images. Currently, a large number of deep learning models emphasize developing deep architectures and ignore capturing the structural features of curvilinear objects, which may lead to unsatisfactory results. Consequently, a new approach that incorporates an ODoS filter as part of a deep learning network is presented to improve the spatial attention of curvilinear objects. Specifically, the input image is transferred into four-channel image constructed by the ODoS filter. In which, the original image is considered the principal part to describe various image appearance and complex background illumination conditions, a multi-step strategy is used to enhance the contrast between curvilinear objects and their surrounding backgrounds, and a vector field is applied to discriminate thin and uneven curvilinear structures. Subsequently, a deep learning framework is employed to extract various structural features for curvilinear object segmentation in medical images. Which can effectively capture the spatial attention of curvature objects. The performance of the computational model is validated in experiments conducted on the publicly available DRIVE, STARE and CHASEDB1 datasets. Compared with ground truths, the presented model acquired high $F_1$ values of 0.826, 0.819 and 0.808, and high accuracy of 0.969, 0.973, and 0.973 on three different datasets, respectively. The experimental results indicate that the presented model yields surprising results compared with those of some state-of-the-art methods.

1. **Introduction**

Curvilinear object segmentation tasks are often encountered in medical image processing and analysis scenarios. The segmentation of such objects is crucial in applications such as the early detection of COVID-19, the screening of retinal fundus disease, and the diagnosis and treatment of lung diseases [1-5]. Compared with generic and unified structure segmentation in medical images, curvilinear object segmentation faces many unique challenges: 1) low contrast between curvilinear objects and their surrounding backgrounds; 2) thin and uneven curvilinear structures; and 3) various complex image representations. Consequently, many methods have been presented to cope with these challenges for curvilinear object segmentation.

Numerous algorithms have been proposed for curvilinear object segmentation in medical images, and they can be divided into three different categories. The first group of classification methods generally designs distinctive handmade features to highlight curvilinear structure representations, such as Hessian matrix-based filters [6,7], tensor-based filters [8,9], stick-based filters [10,11], dynamic evolution models [12,13], active contour models [14,15], and level sets [16]. The second strategy mainly uses deep learning methods to isolate and detect thin curved-lines in medical images under different frameworks, such as UNet network [17], UNet++ network [18], channel and spatial attention networks [19,20], teacher-student networks [21,22], the transformer method and its variants [23,24], SemiCurv [25], and CS2Net [26]. The third category typically combines deep learning networks and handmade features to preserve the inherent curvilinear features of curvilinear structures and achieve improved segmentation accuracy, such methods include D-GaussianNet [27], the local intensity order transformation (LIOT) [28], and the combination of a CNN-based method and a geometry method [29]. Although plausible results have been acquired by using different algorithms have been published, segmentation of weak curvilinear objects remains a formidable task in medical images.

Based on the observation that curvilinear objects have unique shape and structure characteristics, a direct strategy is to detect thin and uneven curvilinear structures by extracting engineered features from medical images. Following this strategy, Li et al. proposed a hardware-oriented method to accurately acquire curvilinear lines based on the Hessian matrix [6], which may lead to discontinuity and fracture of curvature objects. Utilizing a different strategy, Xiao et al. presented a derivative of stick filter consisting of three parallel and gapped sticks to highlight curvilinear structure representation [30], but the distinctive approach applied only the magnitude information, ignoring orientation information. To cope with this problem, Peng et al. designed a unique strategy by taking advantage of orientation and magnitude information to distinguish between curvilinear objects and other structures [31], the approach has great difficulties handling deformed and disrupted curvilinear structures. Recently, Liu et al. applied geodesic models to alleviate the shortcut combination problems encountered in curvilinear structure extraction cases [32], but the approach is only useful in two dimensional space. However, traditional handmade feature extraction algorithms only extract a few features and cannot accurately detect curvilinear structures.

To compensate for the defects of traditional methods, a large number of deep learning models have been proposed to segment curvilinear objects and they have achieved significant improvements over previously developed methods in medical images. For example, Ma et al. presented a cascaded deep neural network based on an anatomically-aware dual-branch structure to achieve curvilinear object segmentation in medical images [33]. Similarly, Roy et al. presented a multi-view deep learning network to detect bright thin curved lines [17]. Later, Li et al. designed an improved IterNet based on structural redundancy to highlight curvilinear structure representations for the diagnosis of retinal vascular diseases[34]. However, their methods based on unet model generally cause blurry edges in segmentation tasks, and fail to accurately acquire the global features in medical images. To improve the detection performance, Barua et al. presented a novel framework to extract multilevel deep features from 18 pre-trained convolutional neural networks [35]. Similarly, Kobat et al. proposed a distinctive method based on horizontal and vertical patch division to capture global and local high-level features for curvilinear structure detection [36]. Especially, their methods [35, 36] can achieve excellent detection results. Utilizing a different method, Li et al. presented a global transformer and a dual local attention network to capture global and local characterizations to achieve satisfactory curvilinear object segmentation results [23]. Similarly, Mou et al. incorporated a self-attention mechanism into the UNet model to learn rich hierarchical representations of curvilinear structures[26]. These transformer model-based methods required a significant computation in training stage. However, deep learning models mainly emphasize finding powerful architectures and ignore capturing the inherent curvilinear features in medical images.

Recently, many distinctive approaches based on deep learning networks and handmade features have been presented to improve the generalizability and segmentation performance achieved for medical images. For instance, Alvarado-Carrillo et al. designed a new technique that combines distorted Gaussian matched filters with adaptive parameters as part of a deep convolutional architecture to detect complicated curvilinear shapes [27]. Using a different method, Shi et al. introduced a novel local intensity order transformation to accurately obtain feature descriptions for curvilinear object segmentation in medical images [28]. Although the above methods can achieve very good results with respect to the segmentation of curvilinear structures in medical images, the detection of weak curvilinear objects remains an enormous challenge.

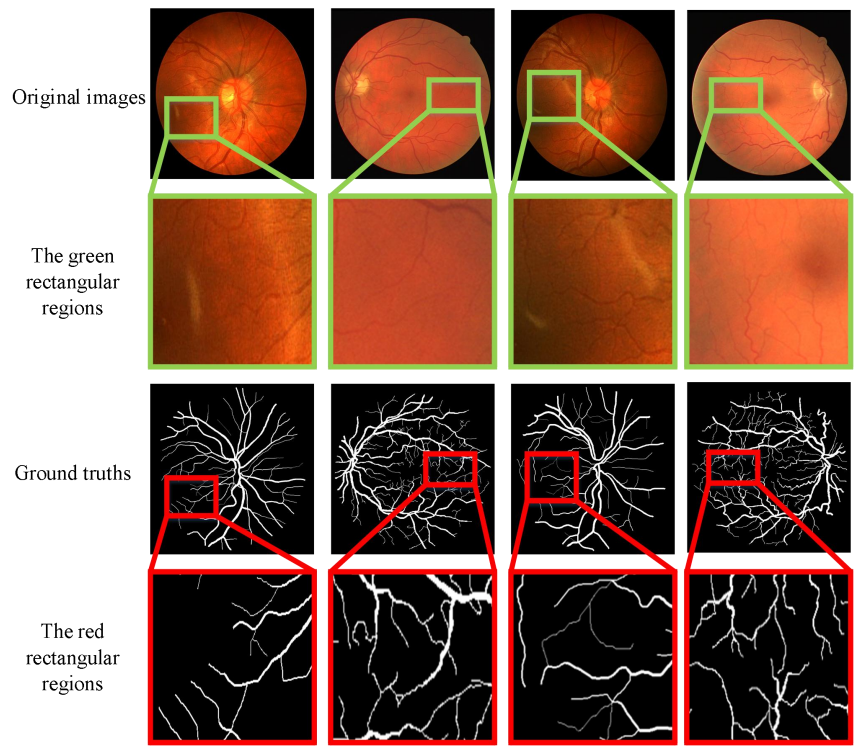

**Fig. 1.**   Weak curvilinear objects.

Weak curvilinear object segmentation is a recognized challenging task in medical image processing field. As shown in Fig. 1, the original images and the corresponding ground truths are given in the first and third rows. The second and fourth rows indicate the corresponding zoomed rectangular regions of the original images and the ground truths, respectively. As marked with green and red rectangular regions, parts of weak curvilinear objects are unable to find and detect from the green rectangular regions due to various factors. Inspired by our previously published methods [10, 30, 31], amplitude information and vector information originated from the ODoS filter are used to highlight weak curvilinear object representation.

What's more, a deep learning model is adopt to achieve a complete curvilinear object segmentation in medical images.

In this study, a valuable and reliable scheme based on an ODoS filter and a deep learning model is presented for weak curvilinear structure segmentation in medical images. The novelties of our model are given as follows. (i) Inspired by previously published works [27, 28], a novel strategy that incorporates the ODoS filter as part of a deep learning network is designed to distinguish between curvilinear structures and undesirable tissues. (ii) The multi-step strategy is employed to enhance the contrast between curvilinear objects and their surrounding backgrounds. (iii) Amplitude information and vector information are combined to describe curvature objects in medical images. (iiii) A deep learning framework is employed to extract various structural features for curvilinear object segmentation in medical images.

The main contributions of our work are as follows:

(i) A novel strategy that incorporates the ODoS filter as part of a deep learning network is presented to overcome different complex issues encountered in curvilinear object segmentation such as various image appearances, low contrast between curvilinear objects and their surrounding backgrounds, thin and uneven curvilinear structures, and complex background illumination conditions.

(ii) The IterNet network is applied to find obscured details of the curvilinear objects from the improved ODoS filter, rather than from the raw input image. The novel strategy has good performance in terms of grasping the spatial information of curved objects in medical images.

(iii) The merits of deep learning networks and handmade features are tightly integrated to generate an efficient image processing application framework for curvilinear object segmentation in medical images.

The rest of this paper is organized as follows. We describe the materials and methods in detail in Section 2, followed by the used of extensive experimental results to verify the validity of our scheme in Section 3. In Section 4, the advantages and disadvantages of different state-of-the-art methods are compared to further illustrate that our scheme can effectively segment curvilinear objects in medical images. Finally, we conclude and give some perspectives in Section 5.

## 2. Materials and Methods

2.1. Datasets and evaluation protocol

To illustrate the validity of the presented method, we use three popular datasets, i.e., DRIVE [37], STARE [38], and CHASEDB1 [39], in our experiments. The DRIVE dataset contains 40 565×584 color retinal images, which are divided into 20 training images and 20 test images. The STARE dataset includes 20 700×605 retinal images, which are split into 10 training images and 10 test images. The CHASEDB1 dataset consists of 28 999×960 images, which are separated into 20 training images and 8 test images.

Following the typical evaluation protocol for curvilinear object segmentation, the true positives (TP), false positives (FP), false negatives (FN), and true negatives (TN) are computed to illustrate the validity of the presented method for curvilinear object segmentation. Then, the classic intersection over Union (IoU), accuracy (ACC), sensitivity (SE), specificity (SP), and F1-score (F1) metrics are used to evaluate the resulting segmentation performance. Mathematically,

$$SE = \frac{TP}{TP+FN} \tag{1}$$

$$SP = \frac{TN}{TN+FP} \tag{2}$$

$$PR = \frac{TP}{TP+FP} \tag{3}$$

$$ACC = \frac{TP+TN}{TP+FP+TN+FN} \tag{4}$$

$$F1 = 2 \cdot \frac{PR \cdot SE}{PR+SE} \tag{5}$$

$$IoU = \frac{TP}{TP+FN+FP} \tag{6}$$

The ACC indicates the proportion of correctly classified positive and negative cases among all participating samples, and F1 represents the similarity between the ground truths and the segmentation results.

## 2.2. Overview of the proposed scheme

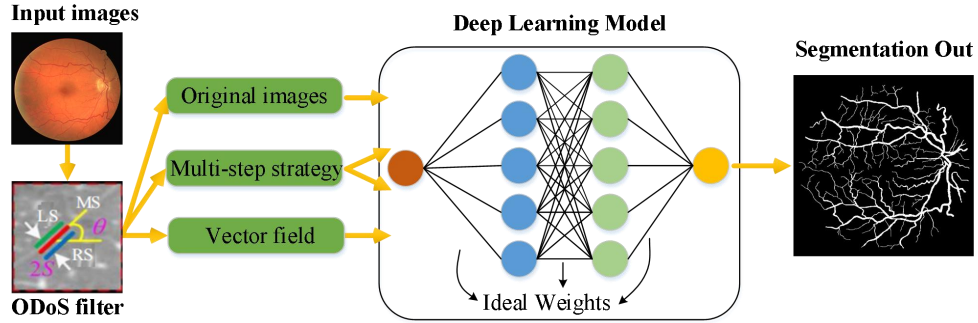

Fig. 2. The framework of the proposed method.

In this study, we present a novel approach based on an ODoS filter and a deep learning network for curvilinear object segmentation in medical images. As shown in Fig. 2, a new approach that incorporates the improved ODoS filter as part of a deep learning network is presented to focus on the spatial attention of curvilinear objects in medical images. Subsequently, a deep learning framework is employed to extract various structural features for curvilinear object segmentation in medical images. The pseudocode is given in Algorithm 1.

---

**Algorithm 1.** Pseudocode of the proposed method

---

**Input**: Image dataset (drd)

**Output**: Precdited_value (pr)

00: **Data processing**

01: Load drd.

02: **for** k=1 to noi **do** //Herein, noi is the number of images in the image dataset

03: Im=drdk; // Read images (Im) from the dataset

04: (v1,v2,v3,v4)=ODoS(Im) // The input image (Im) is transfered into four-channel image (v1, v2, v3, v4) constructed by the ODoS filter

05: Data augmentation and croped the augmented images into 128×128

06: **end for k**

07: **Train stage**

08: **for** epoch=1 to mm **do** // mm is the number of epochs

09:      model.train(v1, v2, v3,v4, GT) // GT represents ground truths, (v1, v2, v3, v4) are the training data

10: **end for epoch**

11: **Test stage**

12: **for** i=1 to nox **do** //Herein, nox is the number of images in the test images

13:     Im=drdi; // Read images (Im) from the test images

14      (v1,v2,v3,v4)=ODoS(Im)

15:      pr=model.test(v1,v2,v3,v4)

16: **end for i**

---

2.3. The improved ODoS filter

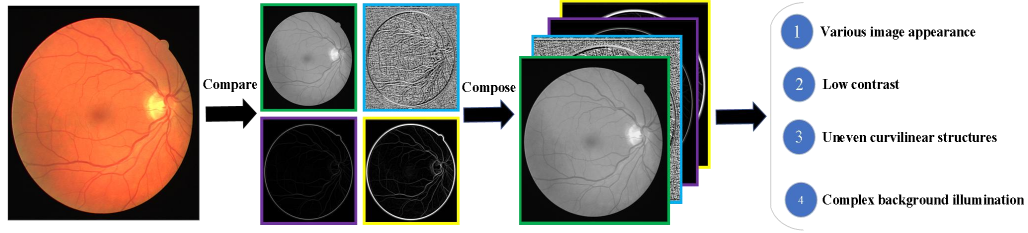

**Fig. 3.** Multi-feature fusion.

In this study, a novel strategy is presented to overcome many challenges in curvilinear object segmentation, such as various image appearances, low contrast between curvilinear objects and their surrounding backgrounds, thin and uneven curvilinear structures, and complex background illumination conditions. As shown in Fig. 2 and Fig. 3, the original image is considered the principal part for describing various image appearances and complex background illumination, the multi-step strategy is used to enhance the contrast between curvilinear objects and their surrounding backgrounds, and the vector field is applied to detect weak, thin and uneven curvilinear structures.

Different medical images have their own unique structures, appearances and backgrounds. To overcome this problem, the original images are applied to describe various image appearances and complex background illumination conditions.

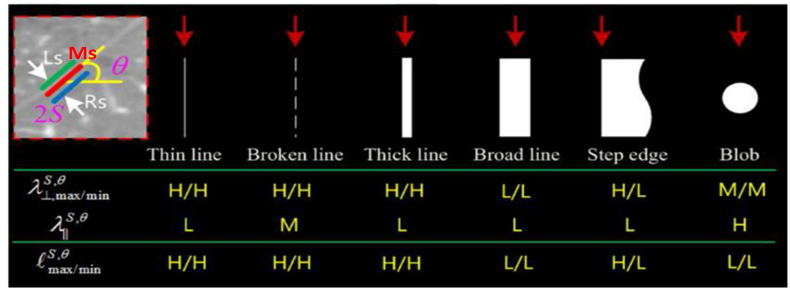

**Fig. 4.** The filtering kernel. (the picture is originates from Xiao et al. [30] with their permission.).

Low contrast between curvilinear structures and their surrounding tissues is another challenge in curvilinear object segmentation. To solve this problem, the multi-step strategy is presented to enhance curvilinear objects and suppress other structures. The idea stems from our previously published works [30, 31], we use three sticks ( left (Ls), middle (Ms), and right (Rs)) to describe the structural differences between curvilinear structures and their surrounding tissues, as shown in Fig. 4. Here, $\theta$ represents the orientation, and $S$ is the interstick spacing. Utilizing $u_L$, $u_M$, and $u_R$ to represent the average intensity levels along the three sticks, two linear enhancement templates are defined as follows [30]

$$\lambda_{\perp,\max}^{S,\theta}(x) = \max(u_M - u_L, u_M - u_R) \tag{7}$$

$$\lambda_{\perp,\min}^{S,\theta}(x) = \min(u_M - u_L, u_M - u_R) \tag{8}$$

Unfortunately, undesired blob shapes are always simultaneously enhanced in medical images. As a result, an intensity standard deviation is applied to solve this problem [30]

$$\lambda_{\parallel}^{S,\theta}(x) = \sqrt{E(I_j^2) - (E(I_j))^2} \tag{9}$$

Here, $E$ and $I_j$ denote the expected value operator and the intensity of the jth pixel along the middle stick, respectively. Therefore, two curvilinear structure measurement functions can be defined as follows [30]:

$$\ell_{\max}^{S,\theta}(x) = \lambda_{\perp,\max}^{S,\theta}(x) - \kappa \cdot \lambda_{\parallel}^{S,\theta}(x) \tag{10}$$

$$\ell_{\min}^{S,\theta}(x) = \lambda_{\perp,\min}^{S,\theta}(x) - \kappa \cdot \lambda_{\parallel}^{S,\theta}(x) \tag{11}$$

Here, $\kappa$ is equal to 0.7 [30].

During the rotation of the filter kernel, the curvilinear structure measurement function values can be changed accordingly. Thus, the corresponding multi-directional integration expressions can be represented with [30]

$$F_{\max}^S(x) = \max(\max_{1 \leq i \leq 2*(L-1)}(\ell_{\max}^{S,\theta_i}), 0) \tag{12}$$

$$F_{\min}^S(x) = \max(\max_{1 \leq i \leq 2*(L-1)}(\ell_{\min}^{S,\theta_i}), 0) \tag{13}$$

Inspired by the application of Hessian matrix decomposition in tubular structure segmentation, a max-min cascaded strategy can be used to suppress the step-edge structures. The associated mathematical expression is as follows [30]:

$$F^S(x) = F_{\max}^S(x) o F_{\min}^S(x) \tag{14}$$

Here, $o$ denotes the cascading operator, which means that $F_{\min}$ is applied after $F_{\max}$ filtering. However, the filtering kernel cannot be directly applied to complex curvilinear structures with different thicknesses. As a result, the multi-step strategy is used to enhance contrast between curvilinear objects and their surrounding backgrounds.

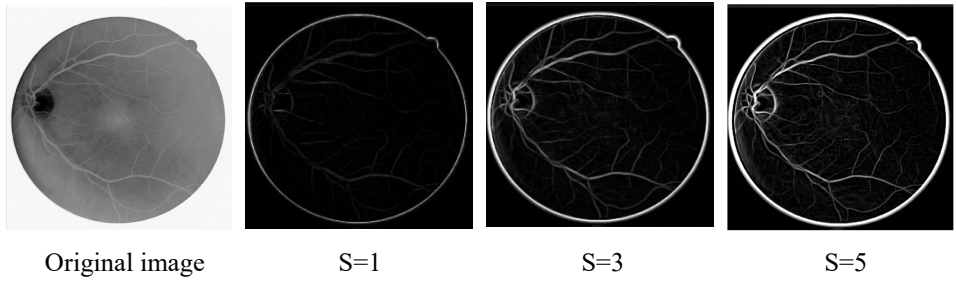

Original image   S=1   S=3   S=5

**Fig. 5.** Curvilinear structure enhancement results $F^S$ obtained with different step size.

As shown in Fig. 5, the step size is too small, and part of the curvilinear structures cannot be detected. When the step size is too large, many undesired shapes cannot be suppressed. Therefore, a multi-step strategy is applied to discriminate curvilinear structures and their surrounding tissues in medical images.

Nevertheless, the greatest challenge is the difficulty of detecting weak, thin and uneven curvilinear structures. Inspired by our previously published methods [10, 30, 31], a vector field is applied to remedy this drawback. The vector representation can be written as [10]

$$\theta^i_{max} = \begin{cases} \arg\max\limits_{1 \le i \le 2(L-1)}(\ell^{S,\theta_i}_{max}), & if\ \ell^{S,\theta_i}_{max} > 0 \\ NaN, & else \end{cases} \quad (15)$$

Equivalently

$$\vec{V}_{max}(\theta^i_{max}) = \begin{cases} (\cos\theta^i_{max}, \sin\theta^i_{max}), & if\ \ell^{S,\theta_i}_{max} > 0 \\ NaN, & else \end{cases} \quad (16)$$

Unfortunately, it is difficult to represent vectors with characteristic symbols in two dimensional images. Based on the fact that the vector $\vec{V}_{max}$ has 2(L-1) different directions, we skillfully use 2(L-1) symbols to indicate different vectors. This approach can not only tactfully represent different vector, but also save space. To investigate the effect of the presented approach, an original image is chosen in Fig. 6(a), and the $\vec{V}_{max}$ vector field is drawn in green arrows in Fig. 6(b). Fig. 6(c) indicates a zoomed rectangular region of Fig. 6(b). For comparison purposes, the transform domain is given in Fig. 6(d). As observed, the transform domain can better represent the direction of curvature structures than vector field. Finally, a distinctive strategy that incorporates the improved ODoS filter as part of a deep learning network is presented to highlight the inspection of shapes, widths, tortuosity levels, and other characteristics for curvilinear object segmentation.

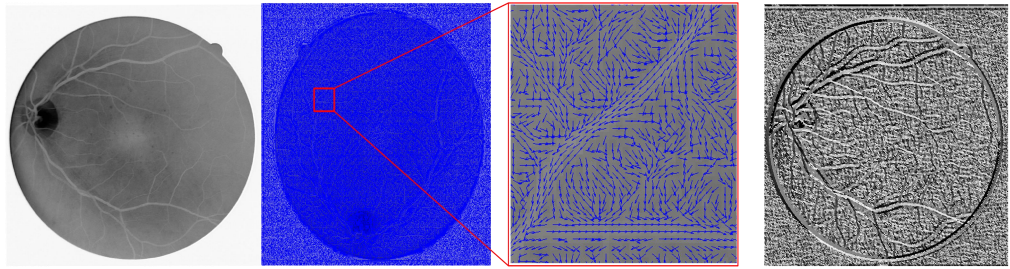

(a).Original image     (b).Vector field     (c).Local vector field     (d).Transform domain

**Fig. 6.** Vector field transform.

## 2.4. Deep learning framework

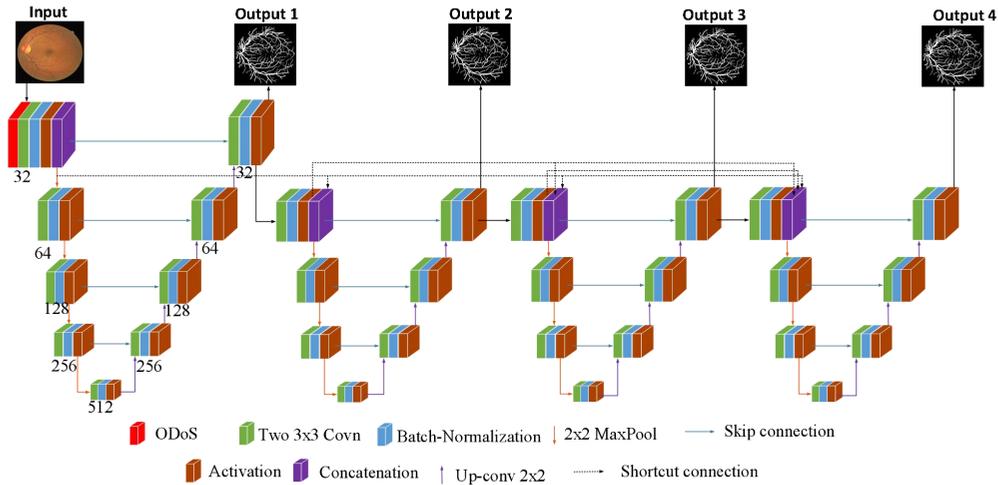

**Fig. 7.** The improved IterNet model

Inspired by previously published works [27, 28, 34], an improved deep learning framework that incorporates the ODoS filter as part of the IterNet is presented. The purpose of this strategy is to alleviate the dependence of traditional deep learning techniques on large datasets, by aggregating robustness through a hybrid design that integrates both a priori knowledge about curvilinear objects and deep learning models. IterNet, which consists of multiple iterations of the mini-UNet uses skip-connections and weight-sharing to facilitate training [34]. Unlike the traditional IterNet model, the first convolution layer channel is changed from three to four channel for four-channel inputs constructed by the ODoS filter. In other words, the improved ODoS filter that considers structural features, placed at the beginning of the presented deep learning framework to illustrate that spatial attention should focus on the curvilinear structures in medical images. In addition, we keep the subsequent layers the same as those of the original IterNet [28, 34], as shown in Fig. 7.

## 3. Experimental results

3.1 Datasets

The presented model is implemented in the PyTorch library with an NVIDIA GeForce RTX 3060 GPU. In which, the overall optimizer is adaptive moment estimation (Adam) in the program, the initial learning rate is set to 0.001 and the maximum number of epochs is equal to 100. To facilitate the observation and objective evaluation of the presented model, we apply our scheme on three widely used datasets: DRIVE [37], STARE [38] and CHASEDB1 [39]. For comparison purposes, five state-of-the-art methods introduced by Zhou et al. [18] Shi et al. [28], Sha et al. [40], Cao et al. [41] and Chen et al. [42] are implemented and applied to the same datasets. Both a visual inspection and a quantitative evaluation illustrate that our scheme has good curvilinear object segmentation performance in medical images.

3.2 Implementation details

In order to compared with recently published methods, we adopt the same data augmentation strategy [28] to increase the training data and avoid over-fitting. The images are randomly rotated from -180 to 180 degrees, and shear from -0.1 to 0.1. Simultaneously, the rotated images undergoes horizontal and vertical flipping operations. Subsequently, these images need to be randomly shifted and zoomed. Finally, the augmented images are croped into 128×128.

It is worth noting that for the sake of fairness, we follow the same training/test division method as the relevant works. Moreover, all models have the same epochs, learning rate and batch size. In which, the epoch is equal to 100, the learning rate is equal to 0.001, and the batch size is equal to 4.

3.3. Visual inspection

To demonstrate the effectiveness of the proposed framework, we apply the model trained on three datasets. As illustrated in Fig. 8, the original images and the corresponding ground truths are given in the first and second rows, respectively. The segmented curvilinear objects obtained with Unet++ [18], Transformer-UNet [40], Swin-UNet [41], TransUNet [42], LIOT [28] and the presented framework are drawn in the third, fourth, fifth, sixth, seventh and eighth rows, respectively. As labeled with blue shapes, the drawback of the comparison methods is that they cannot completely detect weak, thin and uneven curvilinear structures. The main reason for this is that these methods focus on developing deep architectures and ignore capturing the structural features of curved objects. In contrast, the presented model can effectively describe the curvilinear structure features by introducing an ODoS filter to classify the curvilinear structures from the background more effectively. Specifically, amplitude information, vector information and multi-step strategy are tightly combined to enhance the contrast between weak linear structures and their surrounding tissues for weak curvilinear object detection.

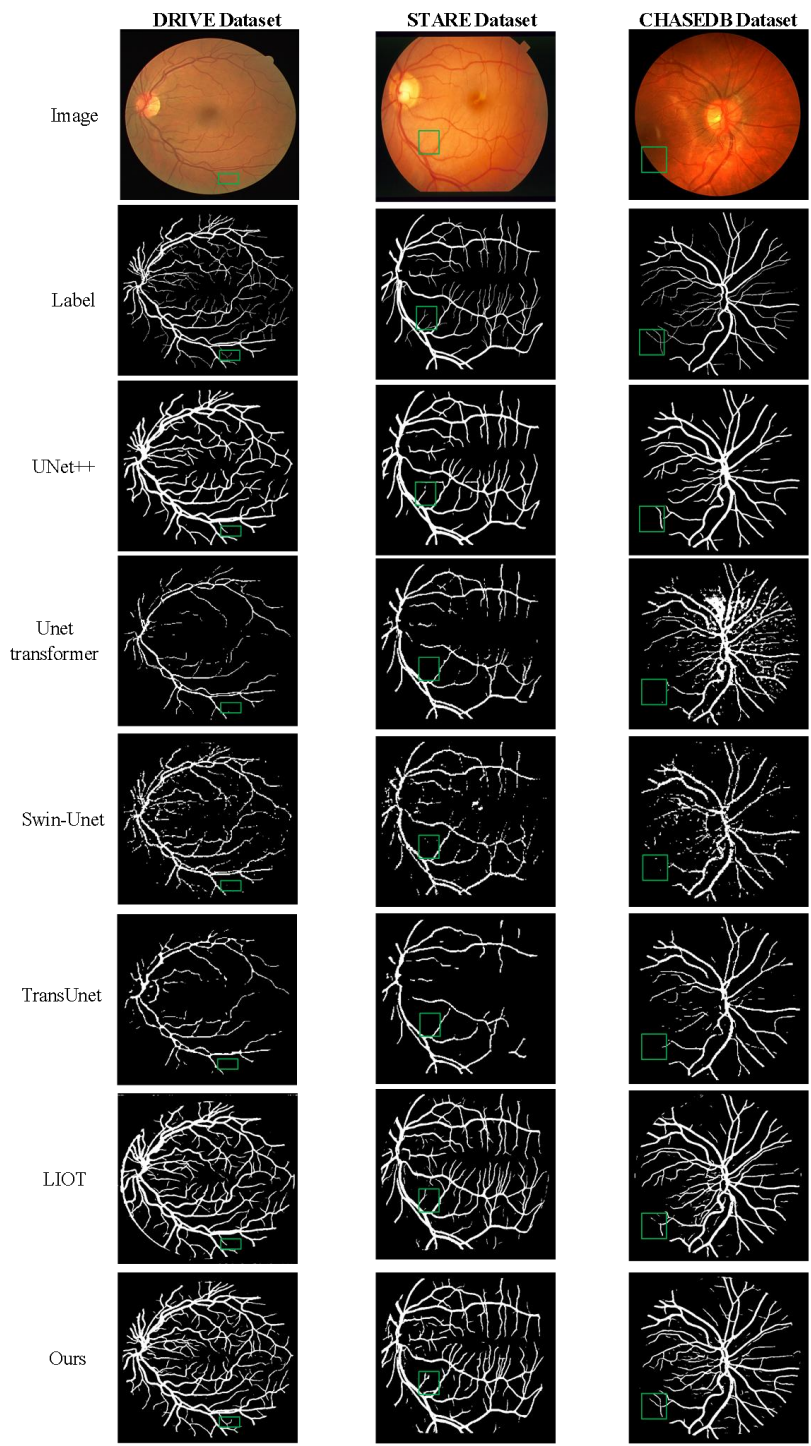

**Fig. 8.** Experimental results obtained with different methods were validated on the DRIVE, STARE, and CHASEDB1 datasets.

## 3.4. Quantitative evaluation

**Table 1. Quantitative evaluation results of different methods.**

| Dataset | Year | Methods | IoU | $F_1$ | ACC | SE | SP |
|---|---|---|---|---|---|---|---|
| DRIVE | 2020 | UNet++ | 0.712 | 0.812 | 0.958 | 0.816 | 0.979 |
| | 2021 | UNet Transformer | 0.402 | 0.557 | 0.922 | 0.406 | 0.999 |
| | 2021 | Swin-UNet | 0.477 | 0.645 | 0.925 | 0.543 | 0.980 |
| | 2021 | TransUNet | 0.577 | 0.731 | 0.940 | 0.635 | 0.986 |
| | 2022 | LIOT | 0.709 | 0.811 | 0.967 | 0.808 | 0.982 |
| | | Ours | **0.740** | **0.826** | **0.969** | **0.828** | 0.983 |
| STARE | 2020 | UNet++ | 0.707 | 0.818 | 0.965 | 0.809 | 0.983 |
| | 2021 | UNet Transformer | 0.387 | 0.556 | 0.888 | 0.389 | 0.999 |
| | 2021 | Swin-UNet | 0.434 | 0.599 | 0.950 | 0.525 | 0.983 |
| | 2021 | TransUNet | 0.440 | 0.609 | 0.902 | 0.461 | 0.990 |
| | 2022 | LIOT | 0.695 | 0.818 | 0.972 | 0.822 | 0.984 |
| | | Ours | **0.714** | **0.819** | **0.973** | **0.823** | 0.987 |
| CHASEDB1 | 2020 | UNet++ | 0.648 | 0.786 | 0.956 | 0.765 | 0.978 |
| | 2021 | UNet Transformer | 0.487 | 0.647 | 0.956 | 0.533 | 0.993 |
| | 2021 | Swin-UNet | 0.442 | 0.610 | 0.952 | 0.605 | 0.976 |
| | 2021 | TransUNet | 0.584 | 0.737 | 0.961 | 0.710 | 0.982 |
| | 2022 | LIOT | 0.682 | 0.791 | 0.971 | 0.810 | 0.982 |
| | | Ours | **0.707** | **0.808** | **0.973** | **0.821** | 0.984 |

The presented method is validated on three publicly available datasets. As shown in Table 1, some high-quality evaluation results are illustrated. Compared with five state-of-the-art methods, the presented method has better IoU, F1-score and accuracy values in most cases. Such good results are attributed to the excellent capturing of curvilinear structure features by the presented model. Both the visual inspection and quantitative evaluation show that the presented deep learning framework can outperform the compared state-of-the-art methods [18, 28, 40-42] in segmentation of weak curvilinear structures.

3.5. Ablation study

**Table II** Ablation study for different inputs with our method on DRIVE dataset

| Method | F1 | ACC | SE | SP |
| --- | --- | --- | --- | --- |
| Original images+IterNet | 0.820 | 0.968 | 0.826 | 0.982 |
| Multi-step strategy+IterNet | 0.815 | 0.968 | 0.815 | 0.983 |
| Vector field+IterNet | 0.808 | 0.967 | 0.803 | 0.982 |
| LIOT+IterNet | 0.811 | 0.967 | 0.808 | 0.982 |
| Original images+Multi-step strategy +Vector field+IterNet | **0.826** | **0.969** | **0.828** | **0.983** |

The purpose of the ablation study is to verify that introducing the ODoS filter into deep learning frameworks can improve the segmentation performance of curvilinear objects. As shown in Table II, if we fix the specific model and change different inputs, these typical evaluation indicators F1, ACC, SE and SP will be changed. As observed, the presented deep learning framework based on the ODoS filter with four channel inputs acquired high values of IoU, F1, ACC, SE and SP on DRIVE dataset.

4. Discussion

In this paper, a unique method based on an ODoS filter and a deep learning network is presented for curvilinear object segmentation in medical images. The presented method has many specific characteristics and advantages. (i) A unique approach that incorporates an improved ODoS filter as part of a deep learning network is presented to focus spatial attention of curvilinear objects in medical images. Many deep learning methods focus on developing deep architectures and do almost nothing to capture the structural features of curvilinear objects, which may lead to unsatisfactory results. (ii) The IterNet model is applied to find obscured details of the curvilinear objects from the improved ODoS filter, rather than from the raw input image. This novel strategy has good performance in terms of grasping the spatial information of curvilinear objects in medical images. (iii) The merits of deep learning networks and handmade features are tightly integrated to generate an efficient image processing application framework for weak curvilinear object segmentation in medical images. (iiii) The presented deep learning framework is expected to preserve the completeness of weak curvilinear object detection while maximally eliminating the unrelated interference. Detection of weak curvilinear objects in medical images is important for clinical diagnoses.

The proposed method is validated in a comparative manner using three publicly available datasets. Both the visual inspection and the quantitative evaluation indicate that the presented approach can outperform several state-of-the-art methods [18, 28, 40-42] in terms of weak object detection. Compared with manually defined ground truths, the presented method has better IoU, F1-score and accuracy values in most cases. The main reason for this is that these comparison methods focus on developing deep architectures and ignore capturing the structural features of curvilinear objects. In contrast, the combination of structure elements and deep learning methods produces better segmentation results in curvilinear object segmentation tasks.

Compared with these state-of-the-art methods [18, 28, 40-42], the presented method appears to more efficiently perform weak curvilinear object detection. This is ascribed to the well-designed combination of an improved ODoS filter and a deep learning model. Nevertheless ,the presented method has some drawbacks. (i) The primary limitation of the presented method relative to these state-of-the-art methods is its longer computational time. In which, the ODoS filter and the deep learning network require a significant amount of time to extract features of curvature objects. (ii) The second limitation is that this method is mainly used to detect curvilinear objects and cannot be used to detect other shapes. The reason is that the presented framework is derived from the ODoS filter and inevitably inherit part of its inherent attributes. (iii) Due to the limited number of images for model training and testing, although data augmentation is used to expand the dataset, the generalization ability of the deep learning model is still restricted. Although the presented method has some disadvantages compared to these state-of-the-art methods, the completeness of weak curvilinear object detection is improved with our scheme.

5. Conclusion

In this paper, we present a distinctive approach based on an ODoS filter and a deep learning network to focus on spatial attention for curvilinear object segmentation in medical images. Motivated by the observation that curvilinear lines and their adjacent interference tissues often have large magnitudes and direction differences, a new method that incorporates the ODoS filter as part of a deep learning network is presented to emphasize the spatial attention paid to curvilinear objects in medical images. Another contribution of our scheme is that it alleviates the dependence of traditional deep learning techniques on large datasets by aggregating robustness through a hybrid design that integrates both a priori knowledge about the curvilinear objects and deep learning models. The performance of the computational model is validated in experiments conducted on the publicly available DRIVE, STARE and CHASEDB1 datasets. Both the visual inspection and quantitative evaluation demonstrated that the presented deep learning framework can outperform the state-of-the-art methods in terms of weak curvilinear object segmentation. In the future, more sophisticated image transformations are worthy of investigation for curvilinear segmentation. In addition, designing more effective deep learning models [43-46] will still be our primary goal.

**Funding.** This research was supported by the Jiangxi Provincial Natural Science Foundation (nos. 20212BAB202007, 20202BAB212004, 20204BCJL23035, 20192ACB21004, 20181BAB202017), the Hunan Provincial Natural Science Foundation (no. 2021JJ30165), the Hunan Special Funds for the Construction of Innovative Province(Huxiang High-level Talent Gathering Project-Innovative talents) (no. 2019RS1072), the Educational Science Research Project of China Institute of communications Education (no. JTYB20-33), the Scientific and Technological Research Project of Education Department in Jiangxi Province (nos. GJJ190356, GJJ210645) and the Science and Technology project of Changsha City (no. kq2001014).

**Data availability.** Data underlying the results presented in this paper are available in Ref. [37], Ref. [38] and Ref. [39].

**Declarations**
**Disclosures.** The authors declare no conflicts of interest.